\documentclass[]{article}
\usepackage{graphicx,subfigure}

\title{Limitations of estimating turbulent convection velocities from PIV}
\date{}

\author{Roeland de Kat\footnote{Aerodynamics and Flight Mechanics Research Group, University of Southampton, Southampton, UK} \footnote{Corresponding author: r.de-kat@soton.ac.uk}, Lian Gan\footnote{Department of Engineering, University of Cambridge, Cambridge, UK}, James R Dawson$^\ddag$, \\Bharathram Ganapathisubramani$^*$}

\begin{document}
\maketitle
\abstract{Abstract  This paper deals with determination of turbulent convection velocities from particle image velocimetry (PIV). Turbulent convection velocities are of interest because they can be used to map temporal information into space. Convection velocity can be defined in several different ways. One approach is to use the phase-spectrum of two signals with a time-separation. Obtaining convection velocity per wavenumber involves determining a spatial spectrum. PIV data is limited in spatial resolution and sample length. The influence of truncation of both spatial resolution and frequency resolution is investigated, as well as the influences of spatial filtering and measurement noise. These issues are investigated by using a synthetic data set obtained by creating velocity-time data with an imposed spectrum. Results from the validation show that, when applying a Hamming window before determining the phase spectrum, there is a usable range of wavenumbers for which convection velocities can be determined. Simulation of flow evolution, movement into and out of the measurement plane, and measurement noise show that these result in a spread in convection velocities using the current approach. Despite this spread, the most probable calculated convection velocity coincides with the imposed convection velocity. Application of the phase-spectral approach to a turbulent boundary layer with $Re_\tau \approx 2700$, shows there is a range of convection velocities and that the most probable convection velocity is equal to the local mean velocity.}

\section{Introduction}

In experiments it is often impossible to obtain full space and time information. Hot-wire data only captures temporal information at a single point, whereas theories about turbulence primarily describe spatial scales. For completely frozen turbulence, TaylorÕs hypothesis \cite{Taylor1938} can be applied to convert spatial information into temporal information and vice versa. However, in practice, completely frozen turbulence rarely exists. 

\begin{figure}
\center{\subfigure{\includegraphics[width=0.49\textwidth]{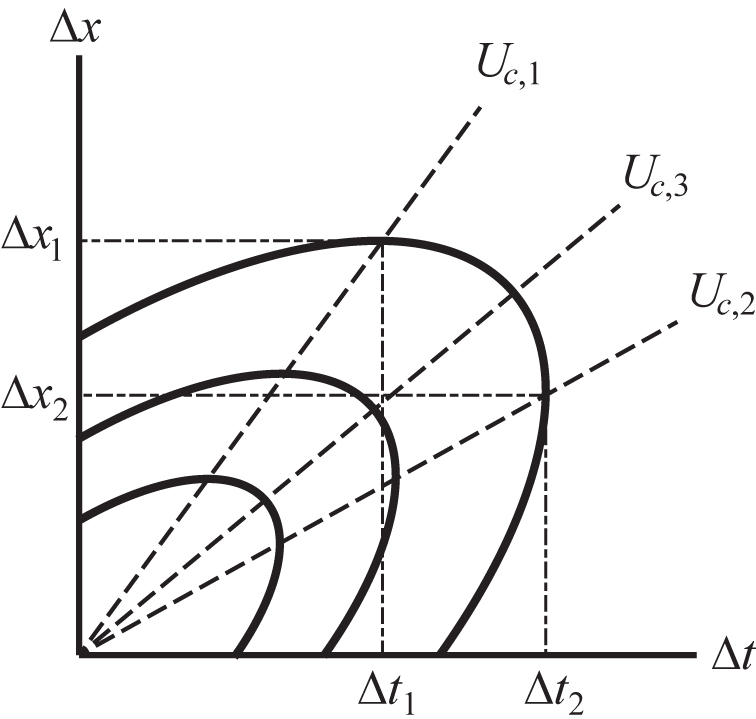}}}
\caption{\label{Velocity} Definitions of convection velocity in the covariance plane (after \cite{Goldschmidt1981})}
\end{figure}

Power spectra and space-time correlations are classical ways to gain insight into the behaviour of turbulent flows \cite{Goldschmidt1981}. Space-time correlation (in space-time or wavenumber-frequency domain) allows determination of convection velocities of turbulent quantities. Its use is well established in combination with hot-wire data (e.g. \cite{Goldschmidt1981}, \cite{Wills1964}, and \cite{Krogstad1998}). 

Convection velocity can be defined in multiple ways (e.g. \cite{Goldschmidt1981}). In the covariance-plane, $\Delta x-\Delta t$, there are at least three possibilities as depicted in figure \ref{Velocity}:
\begin{enumerate}
\item For a given spatial separation ($\Delta x$) find the temporal separation ($\Delta t$) for which the covariance reaches a peak.
\item For a given temporal separation find the spatial separation for which the covariance reaches a peak.
\item Find the ratio of space with time for which the covariance reaches a peak.
\end{enumerate}

For a completely frozen turbulence these definitions are equivalent. For ÔrealÕ turbulence, the third is closest to TaylorÕs hypothesis, since it constitutes a direct relationship between space and time. 

From these spatial and temporal separations the convection velocity can be determined using $u_c = \Delta x/\Delta t$. These three approaches can also be used in the spectral plane, $k_x-\omega$, where a peak in power spectral density is used. The convection velocity can then be determined using $u_c = - \omega/k_x$. Using the different definitions convection velocity can be expressed as a function of time or space separation, frequency or wavenumber.

Furthermore, the possible influence of scale dependent convection velocity makes determining a convection velocity more difficult. Recently, better agreement between experiments and computations was found after correcting for scale dependent convection velocities (see \cite{delAlamo2009} and \cite{Monty2009}).

Del \'Alamo and Jim\'enez \cite{delAlamo2009} have proposed an approach based on phase velocity of the turbulent fluctuations to determine the convection velocity. They showed that a weighted average of this convection velocity is equivalent to a weighted average in the $k_x-\omega$-plane.

In an effort to capture both space and time simultaneously, LeHew et al. \cite{LeHew2011} use time-resolved particle image velocimetry (PIV) to measure a domain of $10\delta \times 5\delta$ in space and $30\delta/U_{local}$ (per individual run, 40 runs in total were taken) in time in a turbulent boundary at $y/\delta = 0.07$. Using space-time correlations they show different convection velocities for different turbulent length scales.

However, in boundary layer flow there are very large-scale coherent streamwise structures with peak in energy for a wavelength of $6\delta$, see Hutchins \& Marusic \cite{Hutchins2007}. This means that LeHew et al. \cite{LeHew2011} capture just over one and a half very large-scale structure in space and approximately five in time.

In addition to the presence of these large scales, the requirements for sufficient spatial resolution \cite{Hutchins2009}, generally results in trade-offs for PIV measurements between spatial resolution and domain size.

Furthermore, to get converged statistics for space-time correlations (in either space-time domain or in wavenumber-frequency domain), not only does the total number of occurrences (of each length/time-scale) need to be enough, also the number of occurrences per block used to determine the correlation or spectrum needs to be sufficiently large. If the number is too low, spectral leakage can influence (all) other scales. It is unclear how this leakage will affect determination of convection velocities.

In view of current limitations in the application of PIV (limited spatial and temporal domains and resolution), this paper will estimate the range of scales for which convection velocity can be determined using limited spatial and/or temporal domains and/or resolutions for each of the different definitions for convection velocity.

\section{Determination of convection velocity}

To determine instantaneous convection velocities, a phase-velocity based approach is applied. This approach determines the convection velocity using the phase difference over a given time separation. This method is closely related to the approach used by del \'Alamo \& Jim\'enez \cite{delAlamo2009}, however, where they take a weighted average, we will use a the different convection velocities to estimate the probability density function ({\em pdf}) of all the instantaneous convection velocities in effort to determine not only the scale dependent influences on convection velocity, but also the range of convection velocities for a specific scale. Resulting {\em pdf}Õs of convection velocities can then be used to determine the mode, mean, and weighted mean of the convection velocity for a particular wavenumber.

We follow a procedure to determine the convection velocity per wavenumber, which is based on the work of Buxton \& Ganapathisubramani \cite{Buxton2011}. The procedure is the following: (i) determine the phase-spectrum, $\Psi(k_x,t)$, (the angle of the cross-spectrum) between the signal at $t$ and $t +\Delta t$; (ii) determine the convection velocity by $u_c(k_x,t) = \Psi(k_x,t)/(2 \pi k_x \Delta t)$, where $k_x=1/\lambda_x$.

Regardless of the method used, the determination of the convection velocity per wavenumber involves determining a spatial spectrum. PIV data is always limited in spatial resolution ({\em WS}) and sample length (FOV), and even if these were not an issue the spatial sample length is limited for the region where the process remains (relatively) stationary. The influence of truncation of both spatial resolution and frequency resolution is investigated, as well as the influences of sampling (overlap factor) and measurement noise. These issues are investigated by using a synthetic data set obtained by creating velocity-time data with an imposed spectrum.

\section{Validation with synthetic data}

Time-space signals are created using PopeÕs spectrum (\cite{Pope2000}, with integral length scale $L$). The spatial signal at the first time step is created by inverse Fourier transform of the imposed spectrum with a random phase. Subsequent time-steps are created by evolving the phase imposing a convection velocity such that $u_c(k_x,t) = U_c = 1$. The ratio between the integral length scale and the smallest length scale is set at $L/\eta = 250$. 

\begin{figure}
\subfigure{\includegraphics[width=0.49\textwidth]{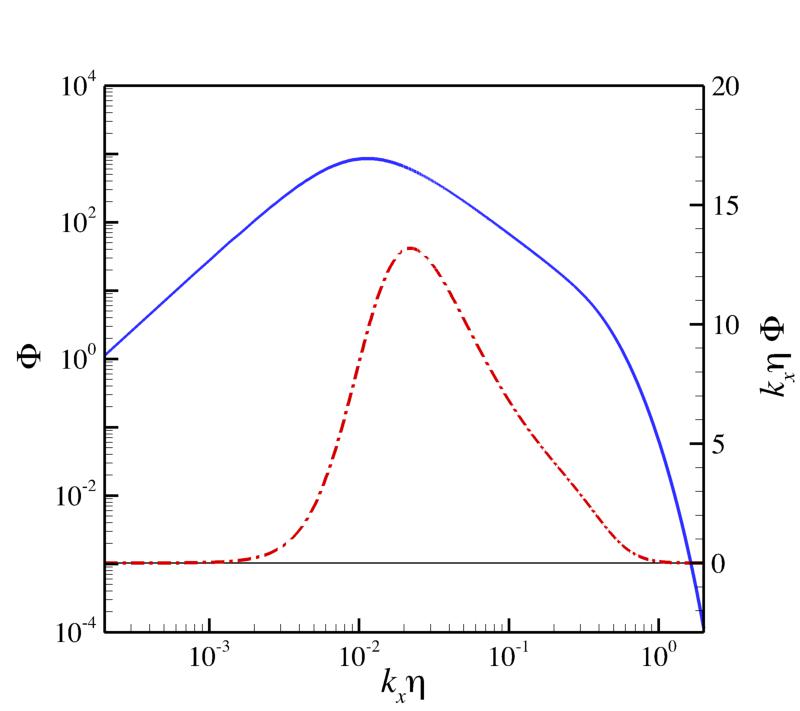}}
\subfigure{\includegraphics[width=0.49\textwidth]{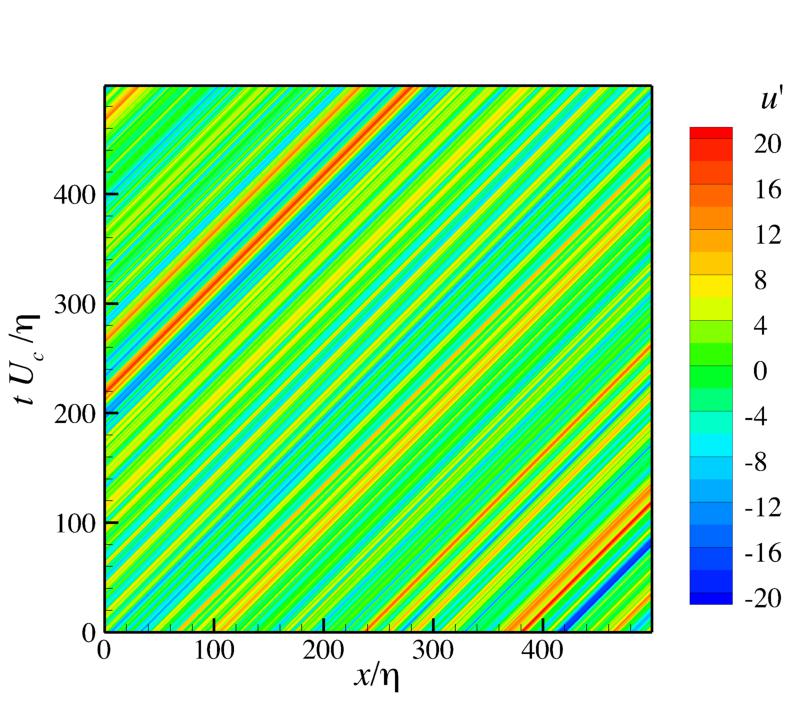}}
\caption{\label{Pope} Synthetic data set creation. {\em Left}: PopeÕs spectrum \cite{Pope2000}. Solid blue line: Power spectrum. Dash-dot red line: Energy spectrum. {\em Right}: Example of a reconstructed space-time plane. One-hundredth of the plane is shown (one-tenth of the spatial and one-tenth of the temporal length).}
\end{figure}

Figure \ref{Pope} shows the power and energy spectrum and an example of the space-time domain. Figure \ref{Pope}{\em left} shows that the reconstruction covers all wavenumbers that have a significant contribution to the energy of the synthetic flow. Figure \ref{Pope}{\em right} shows an example of a reconstructed space-time plane. Regions with equal velocity are inclined at $45^\circ$ showing the convection velocity being equal to unity. This space-time data is subsequently filtered with a spatial filter with filter length $\lambda_f = 5 \eta$ to mimic the low-pass behavior of PIV. 

It is found that to avoid spectral leakage due to the truncation at the sample edges, a Hamming window needs to be applied before taking the Fourier transform. Different windowing functions consistently performed worse (e.g. rectangular window, Hanning window, Blackmann window). All further analysis applies Hamming windows before applying (discrete) Fourier transforms.

\begin{figure}
\subfigure{\includegraphics[width=0.49\textwidth]{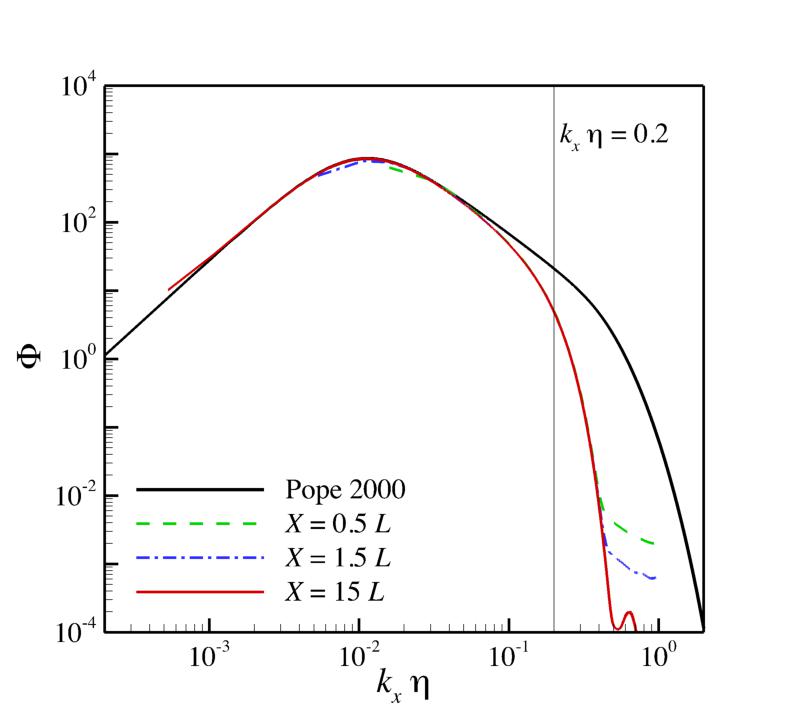}}
\subfigure{\includegraphics[width=0.49\textwidth]{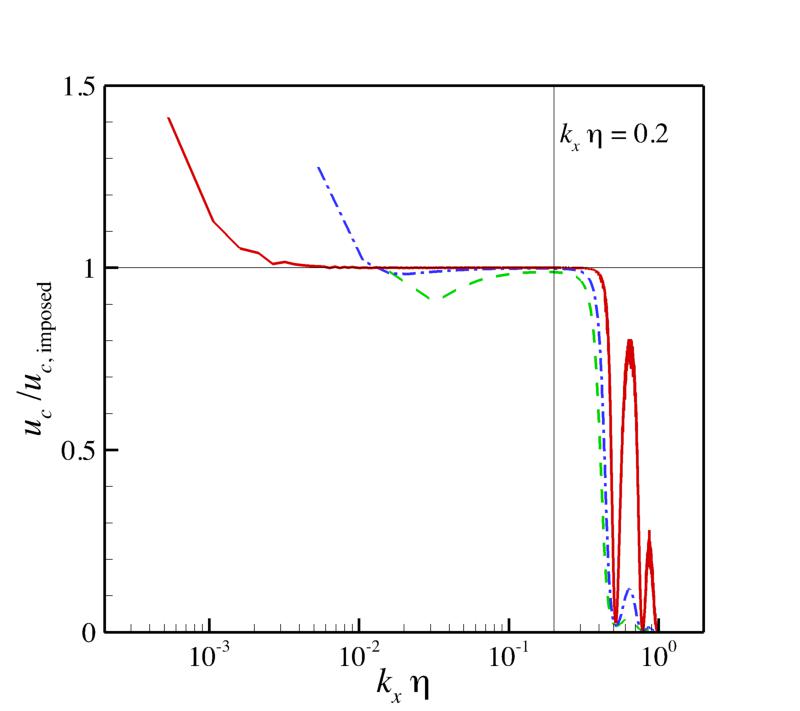}}
\caption{\label{Power} Power spectra and convection velocity based on phase spectra for different domain lengths. The space-time domain was filtered in space with $\lambda_f = 5 \eta$. {\em Left}: Power spectra. {\em Right}: Convection velocity.}
\end{figure} 

Figure \ref{Power} shows results for power spectra obtained with different sample lengths. The power spectra show that the truncation of the domain size, $X$, does not have a severe impact on the estimation of the power spectrum for small wavenumbers. The power spectra at high wavenumber show a distinct deviation from the imposed spectrum. This deviation is an effect of spatial filtering akin to PIV filtering. Foucaut et al. \cite{Foucaut2004} show similar differences for high wavenumbers comparing temporal hot-wire spectra and spatial PIV spectra using the mean velocity to map the hot-wire spectrum into a spatial spectrum.

In contrast to the power spectrum, which is affected for wavenumbers below the filter wavenumber (as indicated in figure \ref{Power}), the phase spectrum is less influenced. The convection velocity based on the phase spectrum (figure \ref{Power}{\em right}) is, therefore, correct for a larger range of wavenumbers than any estimates based on the power spectrum will be. The convection velocity shows no deviation from the imposed convection velocity for wavenumbers larger than $8/X$ and wavenumbers smaller than $1/ \lambda_f$. Synthetic data tests on unfiltered data show that convection velocity can be determined for a wavenumbers $k_x < 1/(4\Delta x)$.

Since the convection velocity is based on the phase spectrum that has a range from $-\pi$ to $\pi$, velocities that cause larger phase angle differences will fold back into this range. Therefore, the convection velocity should be within the range $-1/(2 k_x \Delta t)$ to $1/(2 k_x \Delta t)$.

To briefly summarize the findings, the convection velocity determination using the phase velocity had no observable difference with the imposed convection velocity for:
1.	$1/L < k_x < 1/ \eta$.
2.	$8/X < k_x < 1/(4\Delta x)$.
3.	$k_x < 1/\lambda_f$.
4.	$-1/(2 k_x \Delta t) < u_c < 1/(2 k_x \Delta t)$.
In each case all criteria should be checked and whichever condition is more restrictive is the limiting condition.

\begin{figure}
\subfigure{\includegraphics[width=0.49\textwidth]{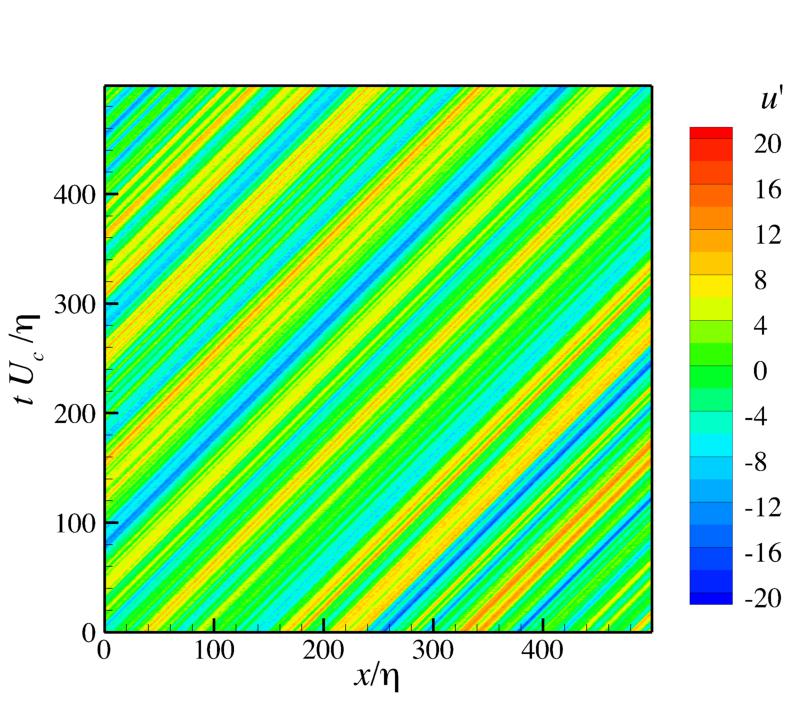}}
\subfigure{\includegraphics[width=0.49\textwidth]{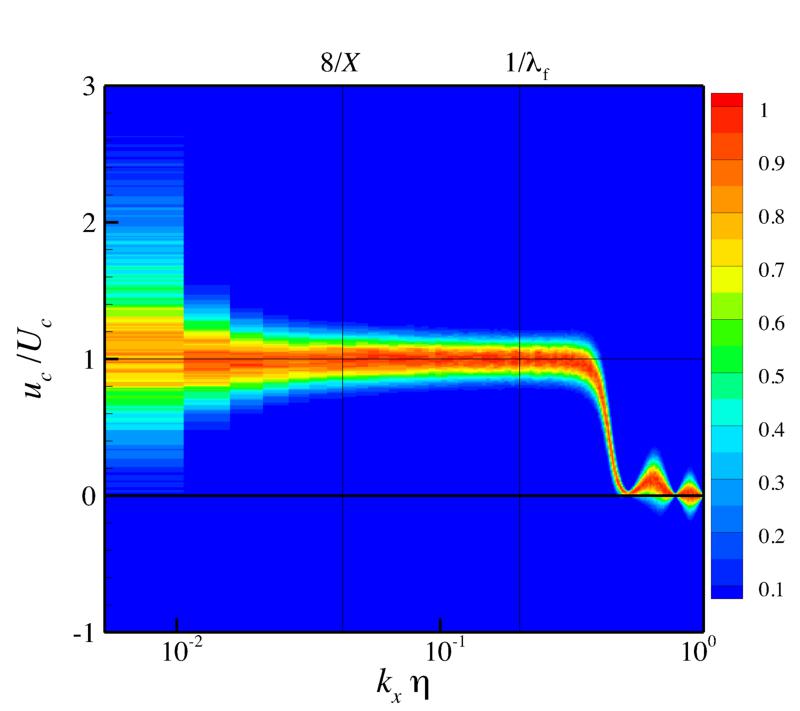}}
\caption{\label{Noise} Addition of synthetic measurement noise with spatial filtering, $\lambda_f = 5 \eta$. The domain size for phase spectra determination is $X = 1.5 L$. {\em Left}: Example of a space-time plane. {\em Right}: {\em pdf} of convection velocity per wavenumber normalized to have the peak of the {\em pdf} equal to 1 for each wavenumber.}
\end{figure} 

To investigate the influence of measurement noise on the determination of convection velocity, random noise fields were created, spatially low-pass filtered to mimic the effect of PIV (see \cite{Foucaut2004}) with the same filter as the signal, i.e. $\lambda_f = 5 \eta$, and added to the signal. Figure \ref{Noise} shows the effect of addition of noise on the determination of convection velocities. Visual inspection of the space-time plane in figure \ref{Noise}{\em left} shows that the effect of noise is minimal, however, the {\em pdf} of convection velocity in figure \ref{Noise}{\em right} shows that there is an appreciable symmetric spread in determined convection velocities around the imposed convection velocity. Nonetheless, the mode of the {\em pdf} coincides with the imposed convection velocity for the range of wavenumbers that satisfy the previously stated restrictions.

\begin{figure}
\subfigure{\includegraphics[width=0.49\textwidth]{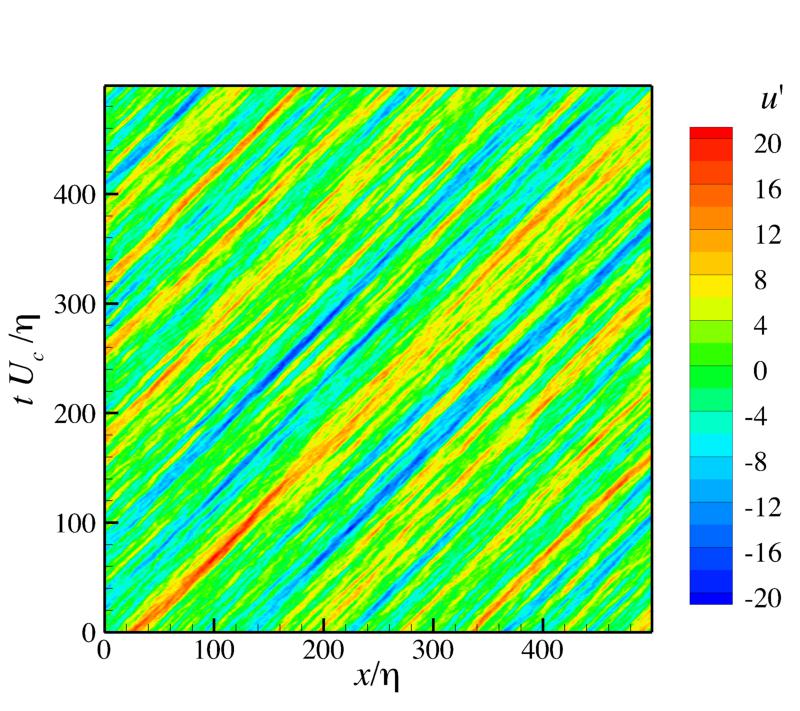}}
\subfigure{\includegraphics[width=0.49\textwidth]{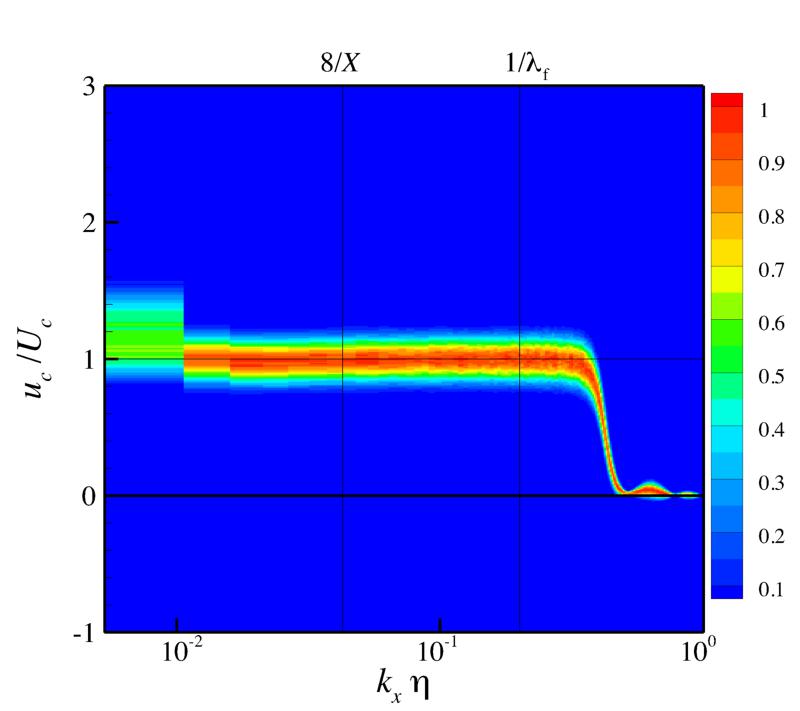}}
\caption{\label{Evolution} Synthetic flow evolution with spatial filtering, $\lambda_f = 5 \eta$. The domain size for phase spectra determination is $X = 1.5 L$. {\em Left}: Example of a space-time plane. {\em Right}: {\em pdf} of convection velocity per wavenumber normalized to have the peak of the {\em pdf} equal to 1 for each wavenumber.}
\end{figure} 

In addition to measurement noise, there are other important factors that should be considered, namely the flow development and movement of the flow into and out of the measurement plane. To mimic these effects a random evolution of the phase angle was added to the phase signal before constructing the space-time plane from PopeÕs spectrum. The resulting temporal correlation between successive steps is 0.95. The space-time plane is shown in figure \ref{Evolution}{\em left} and the corresponding {\em pdf} of convection velocity in figure \ref{Evolution}{\em right}. Even though the space-time plane has a distinctly different appearance than the space-time plane in figure \ref{Noise}, the effect on the determined convection velocities is similar for the range that complies with the previously stated restrictions. The {\em pdf} shows that its mode remains at the imposed convection velocity with a symmetric spread of convection velocities around this mode. 

This shows that flow evolution, movement into and out of the measurement plane, and measurement noise show up as a spread in convection velocities using the current approach. Despite this spread the mode of the resulting {\em pdf}Õs of convection velocity coincides with the imposed convection velocity for the range that complies with the previously mentioned restrictions. The spread will increase with increasing measurement noise, flow development, and/or flow into and out of the measurement plane. However, the mode will remain unaffected.

Now that the validity of the approach has been established, we turn our attention to the turbulent boundary layer and determine convection velocities in it.

\section{Application to turbulent boundary layer data}

Time resolved PIV experiments are performed in a stream-wise wall-normal plane in a turbulent boundary layer in the water tunnel at Cambridge University Engineering Department. The flow is tripped with a glass rod at the beginning of the test-section and PIV measurements are performed 4-5 m downstream of this trip. The nominal flow conditions at the measurement location are $U \approx 0.65$ m s$^{-1}$, $\delta \approx$ 0.1 m, $U_\tau\approx$ 0.027 m s$^{-1}$, and $Re_\tau\approx$ 2700. Two PIV experiments with different field-of-view (FOV) were performed. Particle image pairs are captured and processed using LaVision software DaVis 7.2. The first experiment has a field-of-view (FOV) that covers an area of $37 \times 4.6$ cm (approximately $4\delta \times 0.5\delta$) with a digital resolution of 10 pixels/mm and a total of 25,000 images are acquired in 5 separate sets at 500 Hz. The second experiment has a FOV that covers an area of $17 \times 4.5$ cm (approximately $2\delta \times 0.5\delta$) with a digital resolution of 22 pixels/mm and a total of 50,000 images are acquired in 10 separate sets at 1 kHz. Both experiments covered 50 s of flow. Images are preprocessed using a min-max normalization. Gaussian weighted correlation starts with an initial window size ({\em WS}) of 64 by 64 pixel and finishes at 16 by 16 pixels with an overlap factor of 50\%. This results in a spatial resolution of $\Delta x^+ = 22$ ($l^+ = 43$, \textit{{\em WS}} = 1.6 mm) with a temporal resolution of $\Delta t^+ = 1.5$ and a spatial resolution of $\Delta x^+ = 10$ ($l^+ = 20$, \textit{{\em WS}} = 0.7 mm) with a temporal resolution of $\Delta t^+ = 0.7$ for the large FOV and small FOV, respectively.

\begin{figure}
\subfigure{\includegraphics[width=\textwidth]{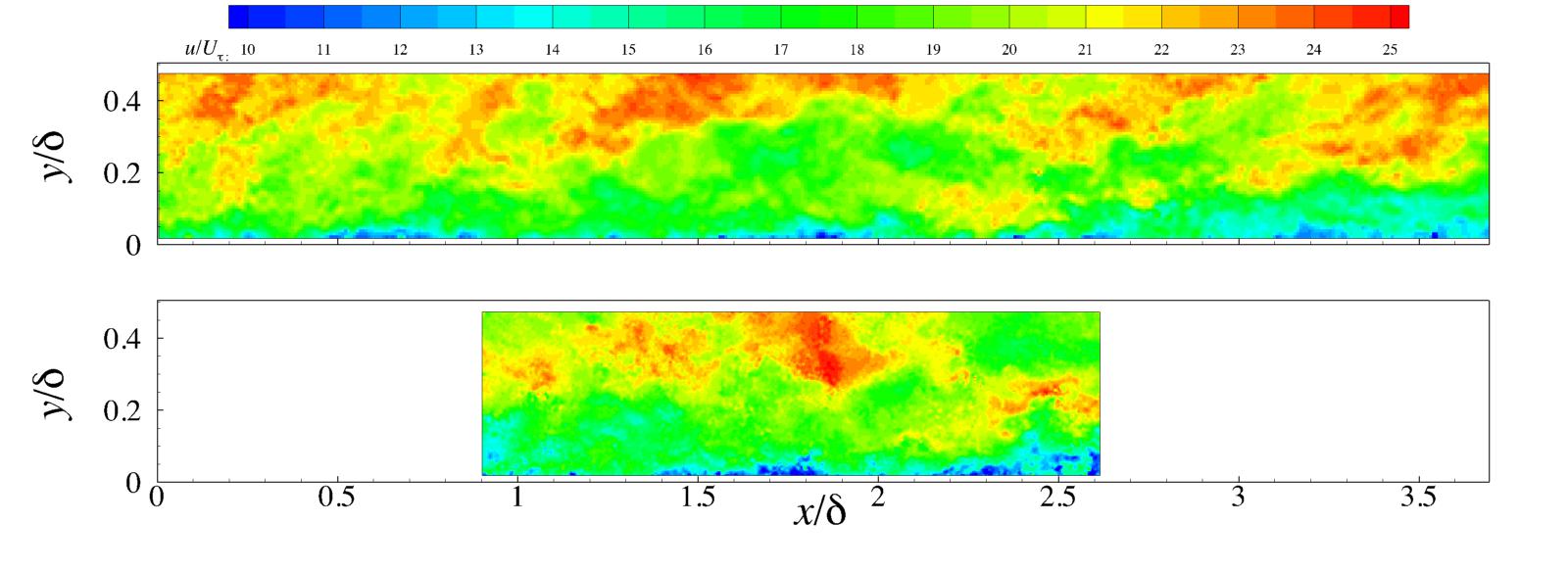}}
\caption{\label{Uexample} Example of instantaneous streamwise velocity results. Note that these are two different experiments. The small FOV is shown in its (approximate) relative location with respect to the large FOV. {\em Top}: Large FOV. {\em Bottom}: Small FOV. }
\end{figure} 

\begin{figure}
\subfigure{\includegraphics[width=0.49\textwidth]{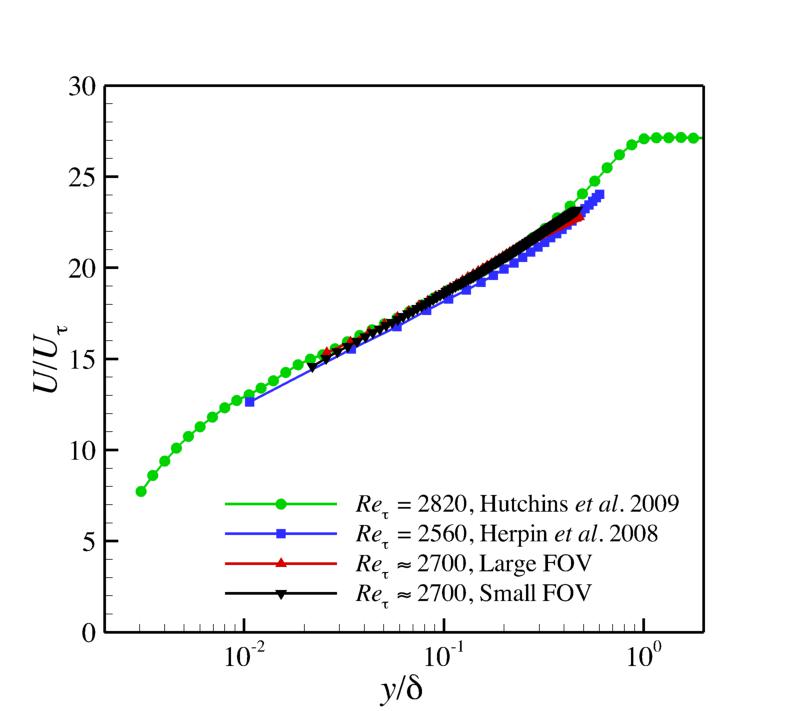}}
\subfigure{\includegraphics[width=0.49\textwidth]{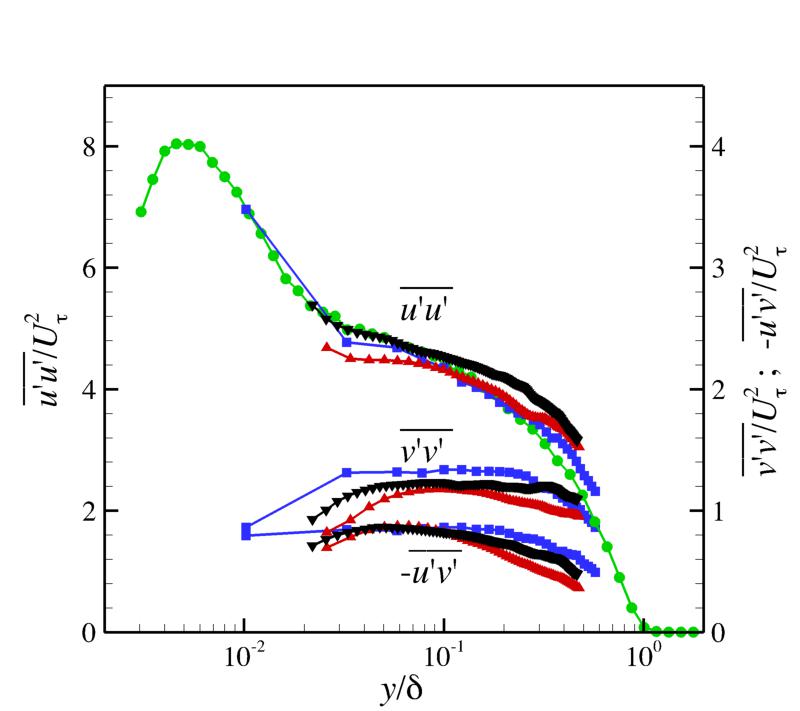}}
\caption{\label{MeanFluct} Mean velocity and turbulent quantities for the current experiment with comparison to other experiments. {\em left}: Mean velocity. {\em right}: Streamwise stress, wall-normal stress, and Reynolds shear stress. {\em Green}: \cite{Hutchins2009} {\em Blue}: \cite{Herpin2008}}
\end{figure} 

Figure \ref{Uexample} show instantaneous snapshots of streamwise velocity for each experiment. Figure \ref{MeanFluct} shows the mean and turbulent fluctuation distributions in the boundary layer, which are in good agreement with previous results at a similar Reynolds number, \cite{Hutchins2009} and \cite{Herpin2008}. The results appear to compare favourable to previous results for similar Reynolds numbers. Data from both sets of experiments are used to calculate the convection velocity using the phase spectral approach.

\begin{figure}
\subfigure{\includegraphics[width=0.49\textwidth]{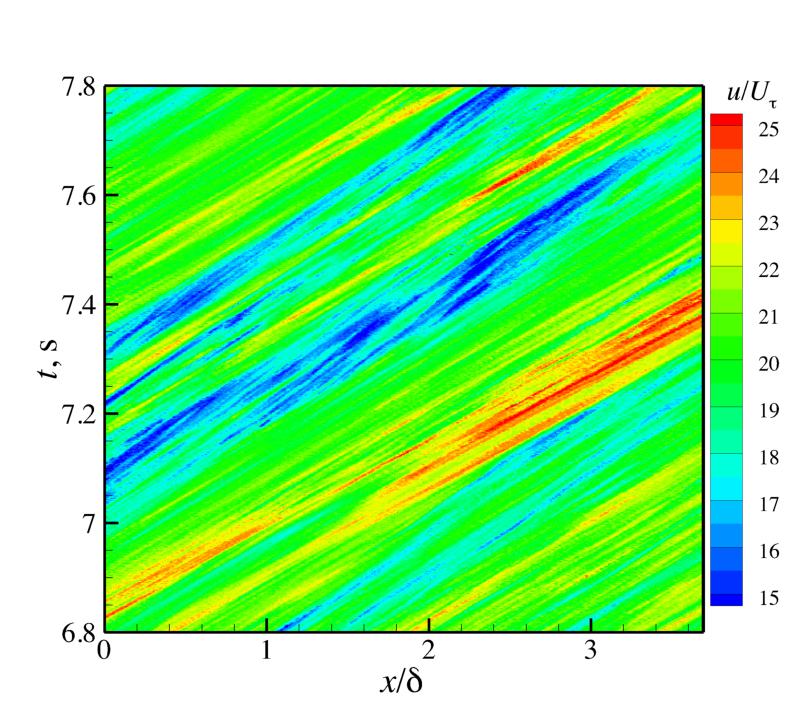}}
\subfigure{\includegraphics[width=0.49\textwidth]{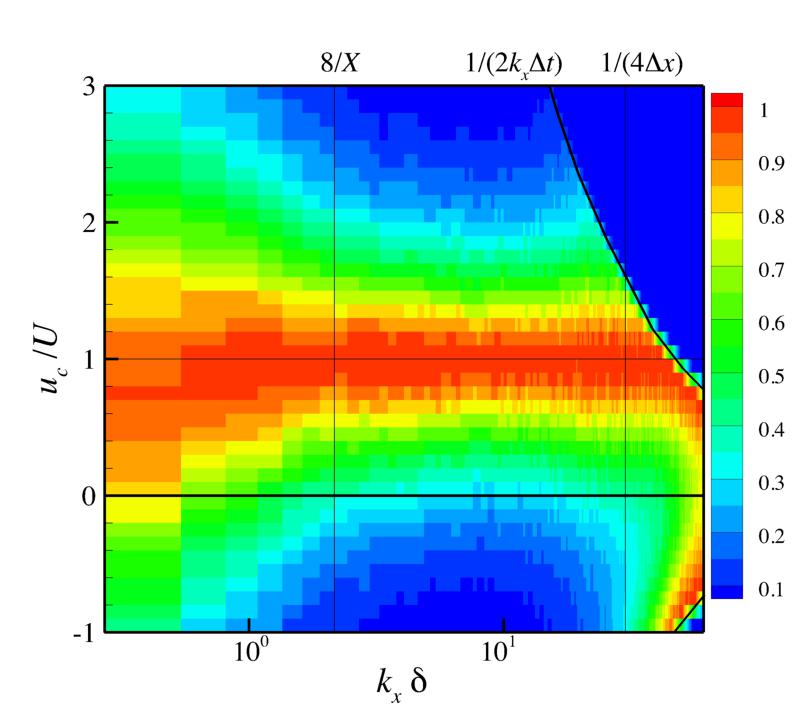}}
\caption{\label{UcLarge} Results from the large FOV, $y/\delta = 0.1$. {\em Left}: Example of space-time plane. {\em Right}: {\em pdf} of convection velocity per wavenumber normalized to have the peak of the {\em pdf} equal to 1 for each wavenumber.}
\end{figure} 

Convection velocity is determined for a wall-normal location in the logarithmic region, $y/\delta = 0.1$, for both the large and the small FOV. An example of a space-time plane obtained together with the {\em pdf} of convection velocity for the large FOV is shown in figure \ref{UcLarge}. The space-time signal, figure \ref{UcLarge}{\em left}, shows elements of noise and flow development (or movement into and out of the plane) that were seen in figure \ref{Noise}{\em left} and figure \ref{Evolution}{\em left}. In addition to these characteristics, there appears to be a range of inclinations of regions of equal velocity, suggesting a range of convection velocities. This is most notable when comparing the relatively high velocity with the relatively low velocity regions. Figure \ref{UcLarge}{\em right} shows the {\em pdf} of convection velocities. There is an appreciable spread in convection velocities, with the mode of the {\em pdf} coinciding with the mean (local) velocity. Whereas it is not possible to indicate what part of the spread in convection velocities is due to measurement noise, flow development, or movement of structures into and out of the measurement plane, the mode shows that the most probable convection velocity is the local mean velocity. This agrees with the findings of Elsinga et al. \cite{Elsinga2012}, who determine convection velocity by tracking hairpin-like vertical structures and found that the mean convection velocity of these structures is equal to the local mean. They found RMS of the spread of convection velocities around the mean convection velocity to be similar to the RMS of the velocity fluctuations around the mean. 

\begin{figure}
\subfigure{\includegraphics[width=0.49\textwidth]{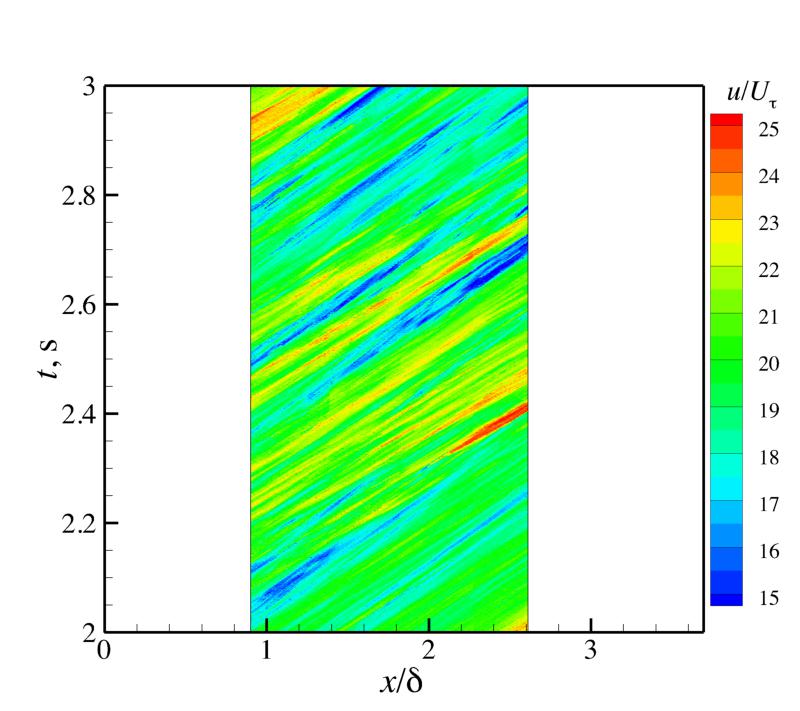}}
\subfigure{\includegraphics[width=0.49\textwidth]{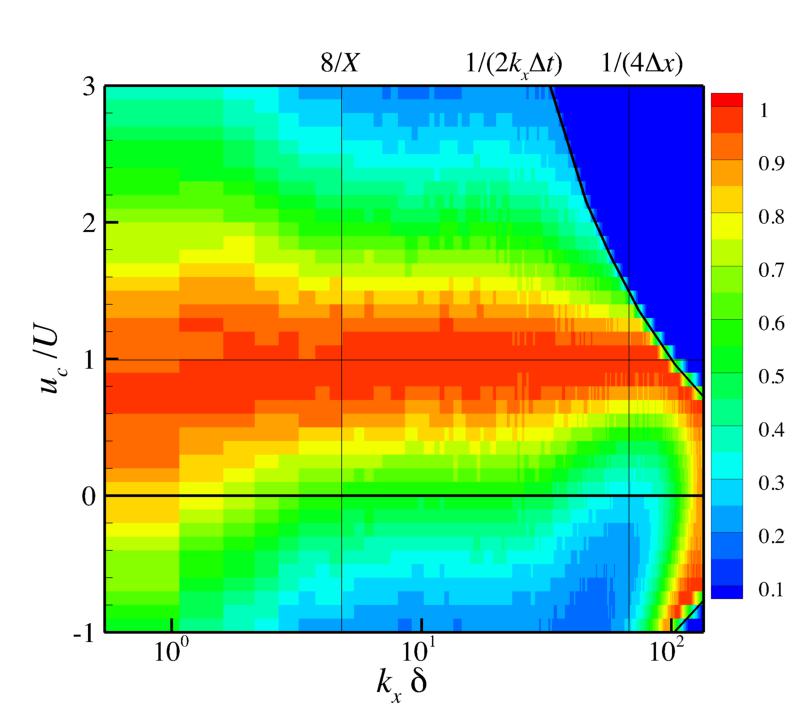}}
\caption{\label{UcSmall} Results from the small FOV, $y/\delta = 0.1$. {\em Left}: Example of space-time plane. {\em Right}: {\em pdf} of convection velocity per wavenumber normalized to have the peak of the {\em pdf} equal to 1 for each wavenumber.}
\end{figure} 

The results for the small FOV, shown in figure \ref{UcSmall}, show a similar trend as the large FOV. There appear to be different inclination of regions of equal velocity in the space-time signal, figure 9{\em left}, and the {\em pdf} of convection velocity shows the mode of the {\em pdf} coincides with the local mean velocity and it has a symmetric spread around the mode. This spread is, however, larger than for the large FOV. Different noise levels and different amplification of noise due to different time-separations can explain this difference in spreads.

\section{Conclusions}

Limitations of determining convection velocity from PIV have been investigated. The determination of convection velocity based on power spectra is severely hampered by the restrictions caused by the low-pass filtering behaviour of PIV. Using an approach based on the phase-spectrum allows us to determine the convection velocity within the region where power spectra based methods will fail.

The approach based on the phase-spectrum has its own limitations. The validation shows that measurement noise and flow evolution have a spreading effect on the convection velocity {\em pdf}, even though they have a distinctly different footprint in the space-time domain. It is therefore difficult to distinguish them from true convection velocity differences based on the {\em pdf} of convection velocity. However, these deviations contribute symmetrically around the true convection velocity and, therefore, the mode of the convection velocity will not be affected. The mode of the {\em pdf} of convection coincides with the imposed velocity for each wavenumber within the range where the convection velocity can be determined.

The turbulent boundary layer results show that the dominant convection velocity in a turbulent boundary layer is the local mean velocity. This agrees with Elsinga et al. \cite{Elsinga2012}, who found that the mean convection velocity of hairpin-like structures is equal to the local mean.

\section*{Acknowledgments}
This work is supported by EPSRC grant EP/I004785/1. The authors would like to thank T.B. Nickels for the inspiration for this work and the use of his water tunnel. Unfortunately, he passed away before the work in this project started.

A version of this work was presented at the \textit{16$^{th}$ Int. Symp. on Applications of Laser Techniques to Fluid Mechanics,} Lisbon, Portugal, 9-12 July 2012.


\begin{thebibliography}{1}

\bibitem{Taylor1938} Taylor, G. I. (1938) ``The spectrum of Turbulence'' {\em Proc. R. Soc. A.}, {\bf 164}(919):476-490
\bibitem{Goldschmidt1981} Goldschmidt, V. W., Young, M. F. \& Ott, E. S. (1981) ``Turbulent convective velocities (broadband and wavenumber dependent) in a plane jet'' {\em J. Fluid Mech.}, {\bf 105}:327-345.
\bibitem{Wills1964} Wills, J. A. B. (1964) ``On convection velocities in turbulent shear flows'' {\em J. Fluid Mech.}, {\bf 20}:417-432.
\bibitem{Krogstad1998} Krogstad, P.-\o A., Kaspersen, J. H. \& Rimestad, S. (1998) ``Convection velocities in a turbulent boundary layer'' {\em Phys. Fluids}, {\bf 10}(4):949-957.
\bibitem{delAlamo2009} del \'Alamo, J. C. \& Jim\'enez, J. (2009) ``Estimation of turbulent convection velocities and corrections to TaylorÕs approximation'' {\em J. Fluid Mech.}, {\bf 640}:5-26.
\bibitem{Monty2009} Monty, J. P. \& Chong, M. S. (2009) ``Turbulent channel flow: comparison of streamwise velocity data from experiments and direct numerical simulation'' {\em J. Fluid Mech.}, {\bf 633}:461-474.
\bibitem{LeHew2011} LeHew, J., Guala, M. \& McKeon, B. J. (2011) ``A study of the three-dimensional spectral energy distribution in a zero pressure gradient turbulent boundary layer'' {\em Exp. Fluids}, {\bf 51}(4):997-1012.
\bibitem{Hutchins2007} Hutchins, N. \& Marusic, I. (2007) ``Evidence of very long meandering features in the logarithmic region of turbulent boundary layers'' {\em J. Fluid Mech.}, {\bf 579}:1-28.
\bibitem{Hutchins2009} Hutchins, N., Nickels, T. B., Marusic, I. \& Chong, M. (2009) ``Hot-wire Spatial resolution issues in wall-bounded turbulence'' {\em J. Fluid Mech.} {\bf 635}:103-136.
\bibitem{Buxton2011} Buxton, O. H. R. \& Ganapathisubramani, B. (2011) ``PIV measurements of convection velocities in a turbulent mixing layer'' {\em J. Phys. Conf. Series}, {\bf 318}:052038.
\bibitem{Pope2000} Pope, S. B. (2000) {\em Turbulent Flows}, Cambridge University Press.
\bibitem{Foucaut2004} Foucaut, M., Carlier, J. \& Stanislas, M. (2004) ``PIV optimization for the study of turbulent flow using spectral analysis'' {\em Meas. Sci. Technol.}, {\bf 15}:1046-1058.
\bibitem{Herpin2008} Herpin, S., Wong, C. Y., Stanislas, M. \& Soria, J. (2008) ``Stereoscopic PIV measurements of a turbulent boundary layer with a large spatial dynamic range'' {\em Exp. Fluids}, {\bf 45}:745-763.

\bibitem{Elsinga2012} Elsinga, G. E., Poelma, C., Schr\"oder, A., Geisler, R., Scarano, F. \& Westerweel, J. (2012) ``Tracking of vortices in a turbulent boundary layer'' {\em J. Fluid Mech.}, {\bf 697}:273-295.


  \end{thebibliography}
\end{document}